# Photo-Seebeck effect in single-crystalline bismuth telluride topological insulator

Anand Nivedan, Arvind Singh, Sandeep Kumar, and Sunil Kumar*
*Department of Physics, Indian Institute of Technology Delhi, Hauz Khas, New Delhi 110016, India*
Email: *kumarsunil@physics.iitd.ac.in

Bismuth telluride is a low energy bulk band-gap topological system with conducting surface states. Besides its very good thermoelectric properties, it also makes a very good candidate for broadband photodetectors. Here, we report temperature-dependent photo-Seebeck effect in a bulk single crystalline bismuth telluride. On light illumination, an electrically biased sample shows distinguishable contributions in the measured current due to both the Seebeck effect and the normal photo-generated carriers within a narrow layer of the sample. Detailed experiments are performed to elucidate the distinction between the Seebeck contribution and the photogenerated current. The temperature-dependence of the photocurrent without Seebeck contribution shows a sign reversal from negative to positive at a specific temperature depending on the wavelength of photoexcitation light.

After the theoretical prediction followed by the experimental realization of the first generation two- and three-dimensional topological insulators [1-3], soon three-dimensional (3d) layered topological insulators became the centre of attention due to many ground breaking discoveries on them. [4,5] Bismuth telluride ($Bi_2Te_3$) is a good example of a 3d topological insulator having its bulk band-gap value in the range of a few hundred meV and conducting Dirac like surface states around the Γ-point in the reciprocal E-k space. [6,7] It is also known to be one of the best thermoelectric material, and this property helps researchers to use it and its alloys as materials in thermoelectric refrigeration. [8,9] Seebeck coefficient is one of the most important criteria to decide the quality of thermoelectric materials. $Bi_2Te_3$ has a very high value of Seebeck coefficient [10] and it shows a transition in its temperature-dependence at a certain temperature below the room temperature.[11]

Anomalous thermoelectricity upon reversing the direction of the temperature gradient was reported in single-crystalline $Bi_2Te_3$ thin films, which is caused due to the coupling between thermoelectric and flexoelectric effects [12]. The later arises due to a stress gradient developed between the $Bi_2Te_3$ film and the substrate used. Owing to the large thermoelectric property of $Bi_2Te_3$, care must be taken while doing optical experiments on this material to avoid any interference of heat-induced Seebeck effect in its optical properties with contributions from bulk and the conducting surface states. [13,14] An increased interest has been seen in this area to either distinguish the interaction of light with metallic surface states and the bulk states or make use of the both for optimised photodetection. [13,15-18]

Photo-induced current generation involves several mechanisms which convert photons into electrical signal. If the material is thermoelectric-semiconducting, the contribution is mainly due to three effects, namely, photo-thermoelectric (Seebeck) effect, photoconductive effect and the effect of photo-induced thermal resistance also known as photo-bolometric effect. [19-21] In case of 3d topological insulators, the contribution in the overall current due to surface states also is not insignificant. [13,14,22] Moreover, such an effect becomes quite prominent at the low-temperatures. While helicity-dependent photocurrent is observed when single-crystalline $Bi_2Se_3$ is illuminated with circularly polarised light [13], the cause for the photocurrent generation under linearly polarized light excitation was proposed due to the topological surface states. [13] Moreover, polycrystalline $Bi_2Te_3$ film was reported in the literature to show negative photocurrent when unpolarized light is incident on its surface and was attributed to light-induced gap opening at the Dirac-point.[14]

In this paper, we report temperature-dependent photo-Seebeck effect in a $Bi_2Te_3$ single crystal. Photoinduced current comprises of the Seebeck contribution as well as the photoconductive part in the bulk and a current due to the metallic surface states. As expected, the magnitude of the photoinduced current depends on the wavelength of the excitation light in the optical region. By proper experimental arrangement, the Seebeck and photocurrent contributions have been estimated and shown at all the sample temperatures.

Freshly exfoliated single-crystalline $Bi_2Te_3$ sample with dimensions of ~5 mm × 3 mm × 0.2 mm was used in the experiments reported here. The crystallinity and the chemical purity were confirmed through X-ray diffraction and Raman spectroscopy, respectively. Figure 1(a) presents an optical image of the nearly clean and uniform surface where one of the boundary regions of the sample can also be seen. Electrical contacts were made in two-point probe geometry for photoinduced current measurements, and in four-point probe geometry for temperature-dependent resistance (R-T) measurements. For all the temperature-dependent measurements reported here, we have used a closed-cycle Helium cryostat system (Janis model SHI-4-2-XG) operating in the temperature range of ~4-475 K. Metallic behavior of $Bi_2Te_3$ as seen in the R-T data presented in Fig. 1(b) implies that the sample has high carrier density which make it difficult for topological surface states (TSS) to be experimentally detected in electrical transport measurements[23]. However, it may be noted that this difficulty can be removed by manipulating the intrinsic defects in TIs, such as by counter doping [6, 24, 25] or electrical gating of the samples [26-28]. Additionally, this problem can also be solved by using very thin samples [29-31] so that the surface to bulk ratio of the conduction increases. Thickness of the $Bi_2Te_3$ sample used in our study is ~200 μm, however, it may be noted that the effective thickness within which the incident light is generating photocarriers is only a few nanometers. The optical penetration depth of 442 nm and 325 nm laser light in $Bi_2Te_3$ is just ~25 nm [32]. In other TIs also like $Bi_2Se_3$, the optical penetration depth in the visible range is of the same order [32, 33]. Larger fraction of the incident laser light interacts with the surface electrons by inducing a band gap in the Dirac's linearly dispersing states [14, 34]. Both of these as well as the Seebeck effect as discussed later in our paper contribute to the experimental observations. Temperature dependent photoinduced conductivity measurements were carried out at 325 and 442 nm wavelengths from He-Cd laser. A source-meter (Keithley, Model 2401) was used for biasing and measuring the current. All the photoconductivity measurements were performed under the application of a very low and fixed dc bias voltage of 0.1 mV.

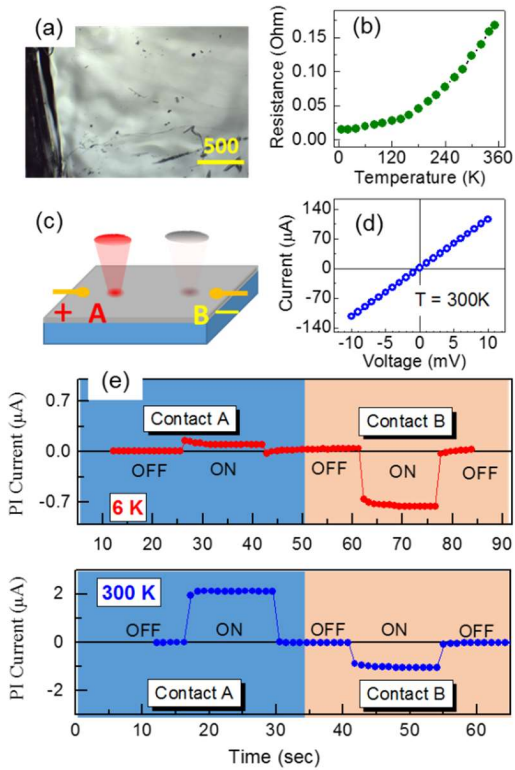

**Fig. 1.** (a) Optical image of the clean and uniform surface of the bulk single-crystalline sample used in present study. (b) Temperature dependent resistance of the sample and (c) experimental geometry used in the photocurrent measurements with light illumination region on the sample being varied between the two contacts of opposite polarity. (d) I-V curve for the sample and (e) the photo-induced current at 6 K and 300 K.

Figure 1(c) depicts the experimental geometry used in the photoinduced (PI) current measurements in this paper. Before doing the actual experiments, current vs voltage (I-V) characteristics were measured, for example, shown in Fig. 1(d) for one case, to ensure that the contacts made were indeed Ohmic in nature. Data were taken for light illuminating the sample near one contact, say contact A or the other contact B (see Fig. 1(c)) ensuring that the distance between the electrical contact and the incident laser spot is same in both the cases. Figure 1(e) shows the values of the photoinduced current corresponding to 1 mW average power light illumination near the two contacts individually taken at 6 K and 300 K in the top and the bottom panels, respectively. Here, the photoinduced current has been corrected by subtracting the current due to minimally applied bias voltage of 0.1 mV. This has been followed throughout the experiments reported in this paper. The most prominent feature in the Fig. 1(e) is the unequal amount of current generation when the light illumination is switched from near contact A to the contact B. At the lowest temperature, i.e., the top panel of Fig 1(e), the value of current generated near contact A is much smaller than that of near contact B. At the same time, nature of the current near contact A is positive whereas it is negative for the near contact B illumination. When the same experiment is performed at the room temperature, i.e., bottom panel in Fig. 1(e), the magnitude of the current near contact A illumination increases while it decreases for the other case. As we will see later in this paper, such an effect arises due to Seebeck contribution in the overall current that changes sign on reversing the direction of the temperature gradient whereas the direction of the photocurrent without Seebeck contribution remains unchanged. It is also important to note that the point of light illumination on the sample near any of the two contacts is kept equidistant from the centre of the sample between the two contacts so that the same value of the thermoelectric Seebeck current is generated in both the cases.

Photo-bolometric effect, i.e., upon light illumination an increase in the resistance with the increasing sample temperature is expected. However, for metallic samples, upon light illumination this effect has an accompanying long relaxation tail in the photocurrent behaviour due to the thermal heating[19, 21]. In contrast, it is clear from the results in Fig. 1(e) that for low intensity laser powers used, the contribution due to photo-bolometric effect in currently studied $Bi_2Te_3$ single crystalline sample is negligible. Therefore, the experimentally measured photoinduced current, i.e., the photo-Seebeck current in Fig. 1(e) is comprised of contributions from the photo-thermoelectric Seebeck effect, the photoconductive effect of the bulk and the metallic surface states. After separating the Seebeck current from the overall photo-Seebeck current, in the below, we have also presented and discussed the temperature dependence of the photocurrent, i.e., current entirely due to the flow of only the photogenerated carriers.

Temperature-dependence of the experimentally measured photo-Seebeck (P-S) current is presented in Fig. 2. The absolute current having positive values for near contact A illumination and negative values for near contact B illumination is seen from Fig. 2(a) for photoexcitation at 442 nm. Clearly, the magnitude of the photo-Seebeck current in each case is highly temperature dependent. Similar results are obtained for light illumination at 325 nm, a summary of both is presented in Fig. 2(b) where the stabilized current values in the light-ON mode is plotted against the sample temperature. It can be seen that for the near contact A illumination, the positive photo-Seebeck current increases with the increasing temperature having a larger variation at around 200 K. On the other hand, for the near contact B illumination, the negative photo-Seebeck current first increases in magnitude until 150 K and beyond 200 K it starts decreasing. It may also be noted from Fig. 2 that the magnitude of the positive photo-Seebeck current is lower than the other at all temperatures below 200 K above which the trend is reversed. These behaviors of the photo-Seebeck current become clear from the discussion later in the paper.

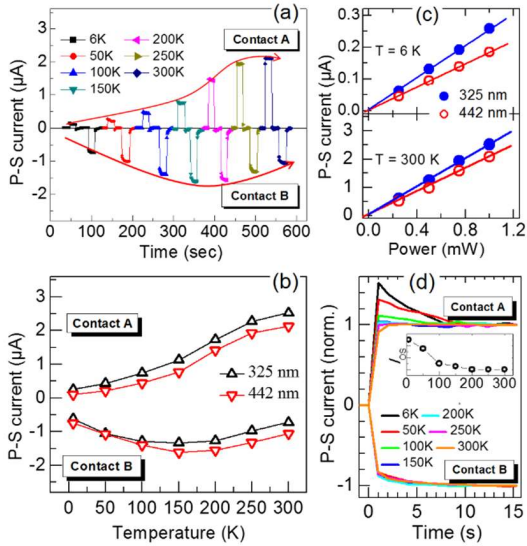

**Fig. 2.** (a) P-S current at 442 nm photoexcitation near contact A and contact B taken at various sample temperatures. (b) Summary of the temperature dependence of the absolute P-S current for near contact A and contact B illumination at the two excitation wavelengths. (c) P-S current as a function of incident laser power at the two excitation wavelengths measured at 6K and 300 K sample temperatures. (d) Normalized photocurrent response as a function of temperature for near contact A and contact B illumination. Magnitude of the overshoot current ($I_{OS}$) as a function of temperature is shown in the inset.

Fig. 2(c) shows the linear variation of the magnitude of photo-Seebeck current with the light intensity at 325 nm and 442 nm for temperatures 6 K and 300 K. Solid straight lines in the figure are guide to the eyes. The trends of these results are preserved for both, near contact A and B illuminations. The linearity in the total photo-Seebeck current is as expected at the light intensity levels used in the current study. [13, 14] At any given sample temperature T inside the cryostat, with the increasing light intensity, an increasing local heating at the point of laser illumination on the sample creates a larger temperature difference locally with respect to the surrounding regions. Therefore, with the increasing light intensity, a characteristic linearly increasing Seebeck current as well as photocurrent will contribute. In Fig. 2(d) we have shown the photo-Seebeck current response normalized with respect to its steady value in the ON state at various sample temperatures for near contact A and contact B illuminations by 442 nm light. Here, for near contact B illumination, the magnitude of the photo-Seebeck current increases rapidly and attains a steady value within a few seconds. The temporal behaviour of the normalized photo-Seebeck current in this case is almost independent of the sample temperature. On the other hand, for near contact A illumination, an overshoot in the photo-Seebeck current can be seen from Fig. 2(d) which eventually settles to a steady value within the time interval of ~10 seconds. The overshoot current disappears at temperatures beyond ~150 K as can be seen from the inset in Fig. 2(d). A possible explanation to this behaviour could be due to accumulation of photogenerated holes near the electrical contact A which creates a hindrance in the movement of photoelectrons towards opposite electrical contact. [35] An exactly opposite behaviour was observed when polarity on the two contacts was reversed thereby confirming the unambiguous results presented above.

As mentioned before, the photo-Seebeck current ($I_{P-S}$) consists of a thermoelectric part, i.e., the Seebeck current ($I_S$) and a photocurrent ($I_P$) due to the bulk as well as the surface states contributions. Following the photoexcitation, $I_S$ and the $I_P$ contribute independently to the resultant $I_{P-S}$. As shown below, our measurements help us clearly distinguish between the $I_P$ and $I_S$ contributions which has not been reported hitherto. This distinction is essential for understanding the topological-thermoelectric effect in $Bi_2Te_3$ under the application of light. Previously, such a distinction between the two effects were ignored by selectively choosing an experimental configuration in which the light is either incident at centre in between the two contacts or by symmetrically illuminating the whole sample. [13, 14]

To separate the Seebeck contribution from $I_{P-S}$, we have performed the experiment in two experimental configurations as depicted schematically in Figs. 3(a) and 3(b) where light illumination takes place alternately near contact A and contact B and the arrows indicate the directions of $I_S$ and $I_P$. Polarities of the contact A (positive) and contact B (negative) are kept same for both the experimental configurations. Here it is obvious that the direction of the $I_P$ cannot be decided initially because it is composed of both, the photoconductive bulk current and the current due to surface electrons. But, the direction of $I_S$ can be determined as clear from the following arguments. There are four possibilities in which the total $I_{P-S}$ can be generated depending upon the sign of $I_S$ and $I_P$. If $I_S$ is considered negative for the light illumination near contact A, then $I_P$ must be positive, and magnitude-wise more than $I_S$ for experimentally observed overall positive photo-Seebeck current ($I_{P-S}$). When the position for laser illumination is switched from near contact A to contact B, $I_S$ would also switch sign and become positive. However, irrespective of the light illumination point between the electrical contacts, $I_P$ will retain its direction as well as magnitude. Irrespective of the position of the laser illumination between the two electrical contacts, the sign and magnitude of the photocurrent $I_P$ remains unchanged for fixed external bias condition. This is due to the fact that it does not depend upon the direction of temperature gradient in the sample. Rather it depends upon the direction of the applied bias. Photogenerated electron-hole pairs within the light penetration depth below the surface of the sample are separated by the applied voltage, which drives electrons towards the positive electrode and holes towards the negative electrode. It is to be noted that the direction of the applied bias has been kept unchanged throughout the experiments. Hence, photocurrent $I_P$ would retain its direction, but the Seebeck current $I_S$ would reverse whenever the position for laser illumination is moved from near one contact to the other. In this case, both the currents, $I_S$ and $I_P$ become positive, thereby, making the total photo-Seebeck current $I_{P-S}$ positive for near contact B illumination. However, this is not what we have experimentally observed. Therefore, the direction of the Seebeck current $I_S$ must always be positive for near contact A illumination and negative for near contact B illumination under the constant external bias condition as shown. Now, assuming the directions of $I_P$ positive in the first experimental configuration in Fig. 3(a), the total $I_{P-S}$ will have an additive contribution from both the $I_S$ and $I_P$. When illumination is changed to near contact B, $I_S$ changes its direction, but the direction of $I_P$ remains the same. Noting that the magnitudes of $I_P$ and $I_S$ remain same upon changing illumination centre, $I_P + I_S$ would be the net photo-Seebeck current when light is incident near contact A. Similarly, $I_P − I_S$ would be the net photo-Seebeck current when light is incident near contact B. These two results have been used to estimate the $I_P$ and $I_S$ separately.

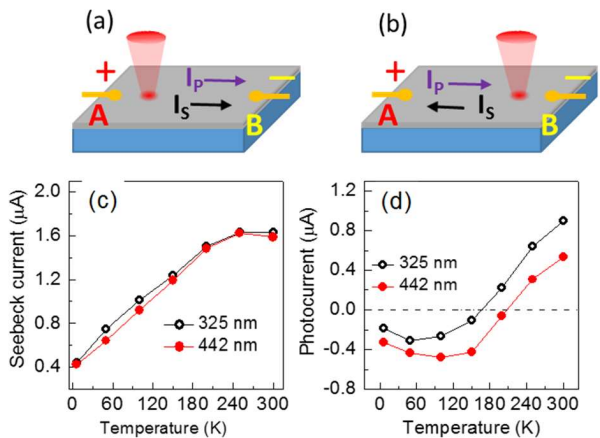

**Fig. 3.** Experimental configurations for the light illumination (a) near contact A and (b) near contact B. The arrows indicate the directions of the photocurrent $I_P$ and the Seebeck current $I_S$. Temperature dependence of (c) the Seebeck current and (d) photocurrent following photoexcitation at 325 and 442 nm wavelengths.

Figures 3 (c) and 3 (d) show the temperature dependence of the extracted Seebeck current and the photocurrent, respectively at 325 nm and 442 nm photoexcitation. The Seebeck current increases linearly with the increasing temperature until it nearly saturates beyond 200 K, independent of the laser wavelength. This kind of transition in the temperature dependence of the Seebeck current is similar to what has been seen for the Seebeck coefficient of $Bi_2Te_3$ where the transition temperature can slightly vary depending on the level of impurities, defects, and other factors in the crystal. [11,36] It is to be noted that the positive value of Seebeck current obtained from our analysis shows that the majority carrier in our sample is holes. In fact, the Seebeck coefficient measurement has been an indirect way to find the majority carriers in materials like $Bi_2Te_3$ and single-layer graphene[12,37]. From Fig. 3 (d) we notice that the photocurrent has negative values at lower temperatures whereas it becomes positive at higher temperatures. The sign reversal in the photocurrent takes place at temperature ~200 K which slightly varies for the two excitation wavelengths.

The photocurrent consists of the contribution from the bulk carriers as well as the surface electrons. Photoconductive effect in the bulk always gives positive value of the current. Hence the negative photocurrent in Fig. 3 (d) is attributed due to the surface electrons. Upon light illumination, depending on the light intensity level, Floquet theory predicts opening of an energy gap in the linear Dirac cone in the density of states of the surface electrons in topological insulators[14, 34]. Such an effect can be significant at low temperatures and in very low bulk band gap bismuth telluride like topological insulators. Therefore, a competition between the contributions of the surface electrons and the bulk carriers can be seen in the resultant temperature-dependent behaviour of the photocurrent.

In conclusion, the photo-Seebeck current analysis in topologically insulating bismuth telluride single crystal has been demonstrated in the paper. By proper experimental arrangements, the pure Seebeck contribution in the photoinduced current can be distinguished from the current due to movement of the photogenerated carriers. The results suggest presence of certain temperature around 200 K at which photocurrent switches sign indicating a discriminating role of the surface states against the bulk at the low temperatures while at high-temperatures their contribution is overwhelmed by the bulk. These results are significant for the understanding of Seebeck effect and the photocarriers generation in bismuth telluride for its potential use in thermoelectric and photodetector applications.

## ACKNOWLEDGMENTS

SK acknowledges the Department of Science and Technology, Government of India for financial support through Science and Engineering Board Project No. ECR/2016/000022. AN acknowledges DST for INSPIRE fellowship.